%% file: RR-7181.tex
\documentclass[a4paper,leqno]{article}

\input{preamble}
\input{command}

\begin{document}

\RRNo{7181}
\makeRR

\input{introduction}
\input{wave}
\input{library}
\input{wave_proof}

\input{conclusion}

\bibliography{biblio}
\bibliographystyle{plain}

\end{document}

%% file: preamble.tex
\usepackage{RRA4}
\usepackage{hyperref}
\usepackage{a4wide}

\RRdate{Janvier 2010}

\RRauthor{
  Sylvie Boldo%
  \thanks[1]{Projet ProVal.
    {\tt \{Sylvie.Boldo,Jean-Christophe.Filliatre,Guillaume.Melquiond\}%
      @inria.fr}.}%
  \thanks[2]{LRI, Universit\'e Paris-Sud, CNRS, Orsay, F-91405}
  \and Fran\c{c}ois Cl\'ement%
  \thanks[3]{Projet Estime. {\tt \{Francois.Clement,Pierre.Weis\}@inria.fr}.}
  \and Jean-Christophe Filli\^atre%
  \thanksref{2}%
  \thanksref{1}
  \and Micaela Mayero%
  \thanks[4]{LIPN, Universit\'e Paris 13, LCR, Villetaneuse, F-93430.\goodbreak
    {\tt Micaela.Mayero@lipn.univ-paris13.fr}.}%
  \thanks[5]{LIP, \'Ecole Normale Sup\'erieure de Lyon, Ar\'enaire,
    Lyon, F-69342}
  \and Guillaume Melquiond%
  \thanksref{1}%
  \thanksref{2}
  \and Pierre Weis%
  \thanksref{3}
}
\authorhead{S. Boldo, F. Cl\'ement, J-C. Filli\^atre, M. Mayero, G. Melquiond,
  \& P. Weis}

\RRtitle{Preuve formelle d'un sch\'ema de r\'esolution de l'\'equation des
  ondes : l'erreur de m\'ethode}
\RRetitle{Formal Proof of a Wave Equation Resolution Scheme: the Method Error}

\RRnote{This research was supported by the ANR projects
  CerPAN (ANR-05-BLAN-0281-04) and
  F$\!\oint$ST (ANR-08-BLAN-0246-01).}

\RRresume{\input{resume}}
\RRabstract{\input{abstract}}

\RRmotcle{\input{motsclefs}}
\RRkeyword{\input{keywords}}

\RRprojets{ProVal et Estime}

\RRdomaine{2 et 5}
\RRtheme{Programmation, v\'erification et preuves\\
  Observation et mod\'elisation pour les sciences de l'environnement}

\RCSaclay

%% file: resume.tex
Les sch\'emas num\'eriques de r\'esolution de l'\'equation des ondes en
dimension~1 sont notoirement convergents.
Nous pr\'esentons une formalisation d\'etaill\'ee du sch\'ema le plus simple et
nous prouvons formellement sa convergence dans le syst\`eme d'aide \`a la
preuve Coq.
La difficult\'e principale se situe dans la d\'efinition ad\'equate des
comportements asymptotiques et du fait que ces comportements ne sont pas
compl\`etement d\'ecrits dans les preuves sur papier.
\`A~notre connaissance, c'est la premi\`ere m\'ecanisation compl\`ete d'une
telle preuve en analyse num\'erique.

%% file: abstract.tex
Popular finite difference numerical schemes for the resolution of the
one-dimensional acoustic wave equation are well-known to be convergent.
We present a comprehensive formalization of the simplest scheme and formally
prove its convergence in Coq.
The main difficulties lie in the proper definition of asymptotic behaviors
and the implicit way they are handled in the mathematical pen-and-paper
proofs.
To our knowledge, this is the first time this kind of mathematical proof is
machine-checked.

%% file: motsclefs.tex
\'equations aux d\'eriv\'ees partielles,
\'equation des ondes acoustiques,
sch\'ema num\'erique,
preuve formelle en Coq

%% file: keywords.tex
partial differential equation,
acoustic wave equation,
numerical scheme,
Coq formal proofs

%% file: command.tex
\usepackage{graphicx}
\usepackage{url}
\usepackage{amsmath}
\usepackage{amsfonts}
\usepackage{color}
\usepackage[utf8]{inputenc}
\usepackage{listings}

\lstdefinelanguage{Coq}{
  comment=[n]{(*}{*)},
  otherkeywords={->, /\\, \\/, =>},
  morekeywords={Goal, Proof, Definition, Theorem, Lemma, Qed, Time, Require, Notation, Variable, Section, End, Fixpoint, forall, exists, fun, let, match, with, end},
  literate={=>}{{$\Rightarrow$}}1 {->}{{$\to$}}1 {/\\}{{$\land$}}1 {<=}{{$\le$}}1 {>=}{{$ge$}}1
}

\newcommand{\derparun}[2]{\frac{\partial {#1}}{\partial {#2}}}
\newcommand{\derpar}[3]{\frac{\partial^{#1} {#2}}{\partial {#3}^{#1}}}
\newcommand{\eqdef}{{\;\stackrel{\text{def}}{=}\;}}
\newcommand{\floor}[1]{\left\lfloor#1\right\rfloor}
\newcommand{\ceiling}[1]{\left\lceil#1\right\rceil}
\newcommand{\Z}{\mathbb{Z}}
\newcommand{\N}{\mathbb{N}}
\newcommand{\R}{{\mathbb R}}

\newcommand{\demi}{\frac{1}{2}}
\newcommand{\prodscal}[3]{\left<#2,#3\right>_{#1}}
\newcommand{\norme}[2]{\left\|#2\right\|_{#1}}
\newcommand{\ps}[2]{\prodscal{}{#1}{#2}}
\newcommand{\n}[1]{\norme{}{#1}}
\newcommand{\nA}[1]{\norme{A(c)}{#1}}
\newcommand{\psdx}[2]{\prodscal{\Delta x}{#1}{#2}}
\newcommand{\ndx}[1]{\norme{\Delta x}{#1}}
\newcommand{\psAh}[2]{\prodscal{A_h(c)}{#1}{#2}}
\newcommand{\nAh}[1]{\norme{A_h(c)}{#1}}
\newcommand{\dotp}[2]{\ensuremath{\langle#1,#2\rangle}}
\newcommand{\norm}[1]{\ensuremath{\|#1\|}}
\newcommand{\FS}[1]{\ensuremath{\mathit{FS}(#1)}}
\newcommand{\xmin}{x_{\rm min}}
\newcommand{\xmax}{x_{\rm max}}
\newcommand{\tmax}{t_{\rm max}}
\newcommand{\jmax}{j_{\rm max}}
\newcommand{\kmax}{k_{\rm max}}
\newcommand{\bfx}{{\bf x}}
\newcommand{\bfdeltax}{{\bf \Delta x}}
\newcommand{\eps}{\varepsilon}

%% file: introduction.tex
\section{Introduction}
\label{sec:introduction} 


Ordinary differential equations (ODE) and partial differential equations
(PDE) are ubiquitous in engineering and scientific computing. They show
up in weather forecast, nuclear simulation, etc.,
and more generally in numerical simulation. Solutions to nontrivial problems are
nonanalytic, hence approximated by numerical schemes over discrete grids.

Numerical analysis is mainly interested in proving the convergence of
these schemes, that is, the approximation quality increases as the
size of the discretization steps decreases. The approximation quality is characterized
by the error defined as the difference between the exact continuous solution and the approximated
discrete solution; this error must tend toward zero in order for the
numerical scheme to be useful.

There is a wide literature on this topic,
{\em e.g.} see~\cite{tho:npd:95,zwi:hde:98}, but no article goes into all the
details. These ``details'' may have been skipped for readability, but they
could also be mandatory details that were omitted due to an oversight. The
purpose of a mechanically-checked proof is to uncover these issues and check
whether they could jeopardize the correctness of the schemes.

This work is a first step toward the development of formal tools for
dealing with the convergence of numerical schemes. It would have been
sensible to start with classical schemes for ODE, such as the Euler
or Runge-Kutta methods. But
we decided to directly validate the feasibility of our approach on the more
complicated
PDE. Moreover, this opens the door to a wide variety of applications, as they
appear in many realistic problems from industry.

We chose the domain of wave propagation because it represents one of the most
common physical phenomena one experiences in everyday life: directly
through sight and hearing, but also via telecommunications, radar, medical
imaging, etc.
Industrial applications range from aeroacoustics to music acoustics (acoustic
waves), from oil prospection to nondestructive testing (elastic waves), from
optics to stealth technology (electromagnetic waves), and even include
stabilization of ships and offshore platforms (surface gravity waves).
We restrained ourselves to the simplest example of wave propagation models,
the acoustic wave equation in a one-dimensional space domain, for it is a
prototype model for all other kinds of wave.
In this case, the equation describes the propagation of pressure variations (or
sound waves) in a fluid medium; it also models the behavior of a vibrating
string.
For simplicity, we only consider homogeneous media, meaning that the
propagation velocity is constant.
Among the wide variety of numerical schemes for approximately solving the 1D
acoustic wave equation, we chose the simplest one: the second order
centered finite difference scheme, also known as the ``three-point scheme''.
Again, for simplicity, we only consider regular grids with constant
discretization steps for time and space.

To our knowledge, this is the first time this kind of mathematical proof is
machine-checked.\footnote{The Coq sources of the formal development are
  available from \url{http://fost.saclay.inria.fr/wave_method_error.php}.}
Few works have been done on formalization 
and proofs on mathematical analysis inside proof assistants, and fewer 
on numerical
analysis. Even real analysis developments are
relatively new. The first developments on real numbers and real
analysis are from the late 90's~\cite{Dut96,Har98,Fle00,May01,GK01}.
Some intuitionist formalizations have been
realized by a team at Nijmegen~\cite{NiqGeu00,Cru02}. Analysis
results are available in provers such as ACL2, Coq, HOL Light,
Isabelle, Mizar, or PVS.
Regarding numerical analysis, we can cite~\cite{May02} which deals,
more precisely, with the formal proof of an automatic differentiation
algorithm. About $\R^n$ and the dot product, an extensive work has been
done by Harrison~\cite{Har05}. About the big~O operator for asymptotic
comparison, a decision procedure has been developed in~\cite{AD07};
unfortunately, we needed a more powerful big O and 
those results were not applicable.

Section~\ref{sec:wave} presents the PDE, the numerical scheme, and their
mathematical properties. Section~\ref{sec:library} describes the basic
blocks of the formalization: dot product, big~O, and Taylor expansions.
Section~\ref{sec:wave_proof} is devoted to the formal proof of the
convergence of the numerical scheme.


%% file: wave.tex
\section{Wave Equation}
\label{sec:wave}

A partial differential equation modeling an evolutionary problem is an
equation involving partial derivatives of an unknown function of several
independent space and time variables.
The uniqueness of the solution is obtained by imposing additional conditions,
typically the value of the function and the value of some of its derivatives at
the initial time.
The right-hand sides of such initial conditions are also called
{\em Cauchy data}, making the whole problem a {\em Cauchy problem}, or an
{\em initial-value problem}.

The mathematical theory is simpler when unbounded domains are
considered~\cite{tho:npd:95}.  When the space domain is bounded, the
computation is simpler, but we have to take reflections at domain
boundaries into account; this models a finite vibrating string fixed
at both ends.  Thanks to the nice property of finite velocity of
propagation of the wave equation, we can build two Cauchy problems,
one bounded and the other one unbounded, that coincide on the domain of
the bounded one.  Thus, we can benefit from the best of both worlds:
the bounded problem makes computation simpler and the unbounded one
avoids handling reflections.
This section, as well as the steps taken at section~\ref{sec:wave_proof} to
conduct the proof of the convergence of the numerical scheme, is
inspired by~\cite{bec:esn:09}.

\subsection{The continuous equation}
\label{sec:continuous}

The chosen PDE models the propagation of waves along an ideal vibrating elastic
string, see~\cite{ach:wpe:73,bg:mcw:94}.
It is obtained from Newton's laws of motion~\cite{new:alm:87}.

The gravity is neglected, hence the string is supposed rectilinear when at
rest.
Let~$u(x,t)$ be the transverse displacement of the point of the string of
abscissa~$x$ at time~$t$ from its equilibrium position.
It is a (signed) scalar.
Let~$c$ be the constant propagation velocity.
It is a positive number that depends on the section and density of the string.
Let~$s(x,t)$ be the external action on the point of abscissa~$x$ at
time~$t$; it is a source term, such that $t=0\Rightarrow s(x,t)=0$.
Finally, let~$u_0(x)$ and~$u_1(x)$ be the initial position and velocity of
the point of abscissa~$x$.
We consider the Cauchy problem ({\em i.e.}, with conditions at $t=0$)
\begin{eqnarray}
  \label{e:L}
  \forall t \ge 0,\; \forall x \in \R, & \quad &
  (L (c) \, u) (x, t) \eqdef
  \derpar{2}{u}{t} (x, t) + A (c) \, u (x, t) =
  s (x, t), \\
  \label{e:L1}
  \forall x \in \R, & \quad &
  (L_1 \, u) (x, 0) \eqdef
  \derparun{u}{t} (x, 0) =
  u_1 (x), \\
  \label{e:L0}
  \forall x \in \R, & \quad &
  (L_0 \, u) (x, 0) \eqdef
  u (x, 0) =
  u_0 (x)
\end{eqnarray}
where the differential operator~$A(c)$ is defined by
\begin{equation}
  \label{e:A}
  A (c) \eqdef - c^2 \derpar{2}{}{x}.
\end{equation}

We admit that under reasonable conditions on the Cauchy data~$u_0$ and~$u_1$
and on the source term~$s$, there exists a unique solution to the Cauchy
problem~(\ref{e:L})--(\ref{e:L0}) for each $c>0$. This is a
mathematical known fact (established for example from d'Alembert's
formula~(\ref{e:dAlembert})), that is left unproved here.

For such a solution $u$, it is natural to associate at each time~$t$ the
positive definite quadratic quantity
\begin{equation}
  \label{e:energiecontinue}
  E (c) (u) (t) \eqdef
  \demi \n{x \mapsto \derparun{u}{t} (x, t)}^2 +
  \demi \nA{x \mapsto u (x, t)}^2
\end{equation}
where $\ps{v}{w}\eqdef\int_\R v(x)w(x)dx$,
$\n{v}^2\eqdef\ps{v}{v}$ and $\nA{v}^2\eqdef\ps{A(c)\,v}{v}$.
The first term is interpreted as the kinetic energy, and the second term as the
potential energy, making~$E$ the mechanical energy of the vibrating string.

This simple partial derivative equation happens to possess an analytical
solution given by the so-called d'Alembert's formula~\cite{dal:rcf:47},
obtained from the method of characteristics~\cite{joh:pde:86},
$\forall t\ge 0$, $\forall x\in\R$,
\begin{multline}
  \label{e:dAlembert}
  u (x, t) =
  \demi (u_0 (x - ct) + u_0 (x + ct)) +
  \frac{1}{2c} \int_{x - ct}^{x + ct} u_1 (y) dy + \\
  \frac{1}{2c} \int_0^t \left(
    \int_{x - c(t - \sigma)}^{x + c(t - \sigma)} s (y, \sigma) dy
  \right) d\sigma.
\end{multline}


One can deduce from formula (\ref{e:dAlembert}) the useful property of finite
velocity of propagation.
Assuming that we are only interested in the resolution of the Cauchy problem
on a compact time interval of the form $[0,\tmax]$ with $\tmax>0$, we suppose
that~$u_0$, $u_1$ and~$s$ have a compact support. Then the property
states that there exists~$\xmin$ and~$\xmax$ with $\xmin<\xmax$ such that the
support of the solution is a subset of
$\Omega\eqdef[\xmin,\xmax]\times[0,\tmax]$.
Furthermore, since the boundaries do not have time to be reached by the
signal, the Cauchy problem set on~$\Omega$ by adding
homogeneous Dirichlet boundary conditions ({\em i.e.} for all $t\in[0,\tmax]$,
$u(\xmin,t)=u(\xmax,t)=0$), admits the same solution.
Hence, we will numerically solve the Cauchy problem on~$\Omega$, but with the
assumption that the spatial boundaries are not reached.

Note that the implementation of the compact spatial domain $[\xmin,\xmax]$
will be abstracted by the notion of finite support (that is to say,
being zero outside of an interval, see Section~\ref{sec:FS})
and will not appear explicitly otherwise.

Note also that most properties of the continuous problem proved unnecessary
in the formalization of the numerical scheme and the proof of its
convergence.
For instance, integration operators and d'Alembert's formula can be avoided as
long as we suppose the existence and regularity of a solution to the PDE and
that this solution has a finite support.

\subsection{The discrete equations}
\label{sec:discrete}

Let $(\Delta x,\Delta t)$ be a point in the interior of~$\Omega$;
define the discretization functions
$j_{\Delta x}(x)\eqdef\floor{\frac{x-\xmin}{\Delta x}}$
and $k_{\Delta t}(t)\eqdef\floor{\frac{t}{\Delta t}}$;
then set $\jmax\eqdef j_{\Delta x}(\xmax)$ and $\kmax\eqdef k_{\Delta t}(\tmax)$.
Now, the compact domain~$\Omega$ is approximated by the regular discrete grid
defined by
\begin{equation}
  \label{e:xjtk}
  \forall k \in [0..\kmax],\,
  \forall j \in [0..\jmax], \quad
  \bfx_j^k \eqdef
  (x_j, t^k) \eqdef
  (\xmin + j \Delta x, k \Delta t).
\end{equation}

Let~$v_h$ be a discrete function over $[0..\jmax]\times[0..\kmax]$.
For all~$k$ in $[0..\kmax]$, we write $v_h^k=(v_j^k)_{0\leq j\leq \jmax}$, then
$v_h=((v_h^k)_{0\leq k\leq \kmax})$.
A function~$v$ defined over~$\Omega$ is approximated at the points of the grid
by the discrete function~$v_h$ defined on $[0..\jmax]\times[0..\kmax]$ by
$v_j^k\eqdef v(\bfx_j^k)$, except for~$u$ where we use the notation 
$\bar{u}_j^k \eqdef u(\bfx_j^k)$ to prevent notation clashes.

\begin{figure}[t]
  \centerline{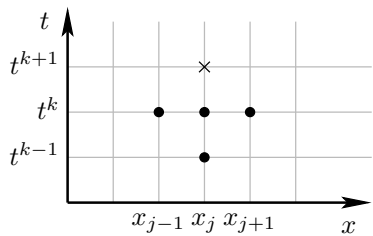}
  \caption{Three-point scheme: $u_j^{k+1}$~($\times$) depends on $u_{j-1}^k$,
    $u_{j}^k$, $u_{j+1}^k$ and $u_{j}^{k-1}$~($\bullet$).}
  \label{fig:scheme}
\end{figure}

Let~$u_{0h}$ and~$u_{1h}$ be two discrete functions over $[0..\jmax]$;
let~$s_h$ be a discrete function over $[0..\jmax]\times[0..\kmax]$.
Then, the discrete function~$u_h$ over $[0..\jmax]\times[0..\kmax]$ is said
to be the solution of the three-point%
\footnote{In the sense ``three spatial points'', for the definition of
  matrix~$A_h(c)$.}
finite difference scheme, as illustrated in Figure~\ref{fig:scheme}, when the
following set of equations holds:
\begin{multline}
  \label{e:Lh}
  \forall k \in [2..\kmax],\,
  \forall j \in [0..\jmax], \\
  (L_h (c) \, u_h)_j^k \eqdef
  \frac{u_j^k - 2 u_j^{k - 1} + u_j^{k - 2}}{\Delta t^2} +
  (A_h (c) \, u_h^{k - 1})_j =
  s_j^{k - 1},
\end{multline}
\begin{eqnarray}
  \label{e:L1h}
  \forall j \in [0..\jmax], & &
  (L_{1h} (c) \, u_h)_j \eqdef
  \frac{u_j^1 - u_j^0}{\Delta t} +
  \frac{\Delta t}{2} (A_h (c) \, u_h^0)_j =
  u_{1,j}, \\
  \label{e:L0h}
  \forall j \in [0..\jmax], & &
  (L_{0h} \, u_h)_j \eqdef
  u_j^0 =
  u_{0,j}, \\
  \label{e:artificial_boundary}
  \forall k \in [0..\kmax[, & &
  u_{-1}^k =
  u_{\jmax+1}^k =
  0
\end{eqnarray}
where the matrix~$A_h(c)$, discrete analog of~$A(c)$, is defined, for
any vector $v_h=((v_j)_{0\leq j\leq \kmax})$, by
\begin{equation}
  \label{e:Ah}
  \forall j \in [0..\jmax], \quad
  \left( A_h (c) \, v_h \right)_j \eqdef
  -c^2 \frac{v_{j+1} - 2 v_j + v_{j-1}}{\Delta x^2}.
\end{equation}

Note that defining~$u_h$ for artificial indexes $j=-1$ and $j=\jmax+1$ is a
trick to make the three-point spatial scheme valid for $j=0$ and $j=\jmax$.

A discrete analog of the energy is also defined by%
\footnote{By convention, the energy is defined between steps $k$
  and $k+1$, thus the notation $k+\demi$.}
\begin{equation}
  \label{e:discreteenergy}
  E_h (c) (u_h) ^ {k+\demi} \eqdef
  \demi \ndx{\frac{u_h^{k+1} - u_h^k}{\Delta t}}^2 +
  \demi \psAh{u_h^k}{u_h^{k+1}}
\end{equation}
where $\psdx{v_h}{w_h}\eqdef\sum_{j=0}^{\jmax} v_jw_j\Delta x$,
$\ndx{v_h}^2\eqdef\psdx{v_h}{v_h}$,\goodbreak
\noindent
and $\psAh{v_h}{w_h}\eqdef\psdx{A_h(c)\,v_h}{w_h}$.

Note that the three-point scheme is parametrized by the discrete Cauchy
data~$u_{0h}$ and~$u_{1h}$, and by the discrete source term~$s_h$.
Of course, when~$u_{0h}$, $u_{1h}$, and~$s_h$ are respectively approximations
of~$u_0$, $u_1$, $f$, then the discrete solution~$u_h$ is an approximation of
the continuous solution~$u$.

\subsection{Convergence}

Let~$\zeta$ and~$\xi$ be in $]0,1[$ with $\zeta\leq 1-\xi$.
The CFL($\zeta,\xi$) condition (for Courant-Friedrichs-Lewy,
see~\cite{cfl:pde:67}) states that the discretization steps satisfy the
relation
\begin{equation}
  \label{eq:cfl}
  \zeta \leq \frac{c \Delta t}{\Delta x} \leq 1 - \xi.
\end{equation}
Note that the lower bound $\zeta$ may seem surprising from a numerical
analysis point of view; the formalization has however shown that it was
mandatory (see Section~\ref{sec:consist}).

The convergence error~$e_h$ measures the distance between the continuous and
discrete solutions.
It is defined by
\begin{equation}
  \label{e:conv_error}
  \forall k \in [0..\kmax],\,
  \forall j \in [0..\jmax], \quad
  e_j^k \eqdef \bar{u}_j^k - u_j^k.
\end{equation}
The truncation error~$\eps_h$ measures at which precision the continuous
solution satisfies the numerical scheme.
It is defined by
\begin{eqnarray}
  \label{e:trunc_error_k}
  \forall k \in [2..\kmax],\,
  \forall j \in [0..\jmax], & &
  \eps_j^{k - 1} \eqdef (L_h (c) \, \bar{u}_h)_j^k - s_j^{k - 1} , \\
  \label{e:trunc_error_1}
  \forall j \in [0..\jmax], & &
  \eps_j^0 \eqdef (L_{1h} (c) \, \bar{u}_h)_j - u_{1,j}, \\
  \label{e:trunc_error_0}
  \forall j \in [0..\jmax], & &
  \eps_j^{- 1} \eqdef (L_{0h} \bar{u}_h)_j - u_{0,j}.
\end{eqnarray}

The numerical scheme is said to be convergent of order~2 if the convergence
error tends toward zero at least as fast as $\Delta x^2+\Delta t^2$ when both
discretization steps tend toward~0.
More precisely, the numerical scheme is said to be convergent of
order~($p$,$q$) uniformly on the interval $[0,\tmax]$ if the convergence
error satisfies (see Section~\ref{sec:o} for the definition of the big O
notation that will be uniform with respect to space and time)
\begin{equation}
  \label{e:convergence}
  \ndx{e_h^{k_{\Delta t}(t)}} =
  O_{[0,\tmax]} (\Delta x^p + \Delta t^q).
\end{equation}

The numerical scheme is said to be consistent with the continuous problem at
order~2 if the truncation error tends toward zero at least as fast as
$\Delta x^2+\Delta t^2$ when the discretization steps tend toward~0.
More precisely, the numerical scheme is said to be consistent with the
continuous problem at order~($p$, $q$) uniformly on interval $[0,\tmax]$ if
the truncation error satisfies
\begin{equation}
  \label{e:consistency}
  \ndx{\eps_h^{k_{\Delta t}(t)}} =
  O_{[0,\tmax]} (\Delta x^p + \Delta t^q).
\end{equation}

The numerical scheme is said to be stable if the discrete solution of the
associated homogeneous problem ({\em i.e.} without any source term, $s(x,t)=0$)
is bounded from above independently of the discretization steps.
More precisely, the numerical scheme is said to be stable uniformly on
interval $[0,\tmax]$ if the discrete solution of the problem without any
source term satisfies
\begin{multline}
  \label{e:stability}
  \exists \alpha, C_1, C_2 > 0,\,
  \forall t \in [0, \tmax],\,
  \forall \Delta x, \Delta t > 0, \quad
  \sqrt{\Delta x^2 + \Delta t^2} < \alpha \Rightarrow \\
  \ndx{u_h^{k_{\Delta t}(t)}} \leq
  (C_1 + C_2 t) (\ndx{u_{0h}} + \nAh{u_{0h}} + \ndx{u_{1h}}).
\end{multline}

The result to be formally proved at section~\ref{sec:wave_proof} states that if
the continuous solution~$u$ is regular enough on~$\Omega$ and if the
discretization steps satisfy the CFL($\zeta$, $\xi$) condition, then the
three-point scheme is convergent of order (2, 2) uniformly on interval
$[0,\tmax]$.

We do not admit (nor prove) the Lax equivalence theorem which stipulates that
for a wide variety of problems and numerical schemes, consistency implies the
equivalence between stability and convergence.
Instead, we establish that consistency and stability implies convergence in
the particular case of the one-dimensional acoustic wave equation.

%% file: scheme.pdf_t
\begin{picture}(0,0)%
\includegraphics{scheme.eps}%
\end{picture}%
\setlength{\unitlength}{4144sp}%
\begingroup\makeatletter\ifx\SetFigFont\undefined%
\gdef\SetFigFont#1#2#3#4#5{%
  \reset@font\fontsize{#1}{#2pt}%
  \fontfamily{#3}\fontseries{#4}\fontshape{#5}%
  \selectfont}%
\fi\endgroup%
\begin{picture}(1972,1427)(3091,-3986)
\put(4906,-3931){\makebox(0,0)[b]{\smash{{\SetFigFont{10}{12.0}{\familydefault}{\mddefault}{\updefault}{\color[rgb]{0,0,0}$x$}%
}}}}
\put(3106,-2716){\makebox(0,0)[b]{\smash{{\SetFigFont{10}{12.0}{\familydefault}{\mddefault}{\updefault}{\color[rgb]{0,0,0}$t$}%
}}}}
\put(4321,-3886){\makebox(0,0)[b]{\smash{{\SetFigFont{10}{12.0}{\familydefault}{\mddefault}{\updefault}{\color[rgb]{0,0,0}$x_{j+1}$}%
}}}}
\put(3781,-3886){\makebox(0,0)[b]{\smash{{\SetFigFont{10}{12.0}{\familydefault}{\mddefault}{\updefault}{\color[rgb]{0,0,0}$x_{j-1}$}%
}}}}
\put(4051,-3886){\makebox(0,0)[b]{\smash{{\SetFigFont{10}{12.0}{\familydefault}{\mddefault}{\updefault}{\color[rgb]{0,0,0}$x_j$}%
}}}}
\put(3196,-2986){\makebox(0,0)[rb]{\smash{{\SetFigFont{10}{12.0}{\familydefault}{\mddefault}{\updefault}{\color[rgb]{0,0,0}$t^{k+1}$}%
}}}}
\put(3196,-3256){\makebox(0,0)[rb]{\smash{{\SetFigFont{10}{12.0}{\familydefault}{\mddefault}{\updefault}{\color[rgb]{0,0,0}$t^{k}$}%
}}}}
\put(3196,-3526){\makebox(0,0)[rb]{\smash{{\SetFigFont{10}{12.0}{\familydefault}{\mddefault}{\updefault}{\color[rgb]{0,0,0}$t^{k-1}$}%
}}}}
\end{picture}%

%% file: library.tex
\section{The Coq Formalization: Basic Blocks}
\label{sec:library}

We decided to use the Coq proof assistant~\cite{Coq}, as Coq was
already used to prove the floating-point error~\cite{Bol09} of this
case study.  All our developments use the Coq real standard
(classical) library.  Numerical equations, numerical schemes,
numerical approximations deal with classical statements, and are not
in the scope of intuitionist theory.

\subsection{Dot product}
\label{sec:dot_prod}

The function space $\Z\rightarrow\R$ can be equipped with pointwise
addition and multiplication by a scalar.
The result is a vector space.
In the following, we are only interested in
functions with finite support, that is the subset
\begin{displaymath}
  F \eqdef \{ f:\Z \rightarrow \R \mid \exists a,b\in\Z, \forall i\in\Z,
  f(i)\not=0 \Rightarrow a\le i\le b \},
\end{displaymath}
which is also a vector space. Then it is possible to define a dot
product on $F$, noted $\dotp{.}{.}$, as follows:
\begin{equation}\label{eq:dotp}
\dotp{f}{g} \eqdef \sum_{i\in\Z}f(i)g(i)
\end{equation}
and the corresponding norm $\norm{f} \eqdef \sqrt{\dotp{f}{f}}$.
The corresponding Coq formalization is not immediate, though.
One could characterize $F$ with a dependent type, but that would make
operation $\dotp{.}{.}$ difficult to use (each time it is applied,
proofs of finite support properties have to be passed as well).
Instead, we define  $\dotp{.}{.}$ on the full function space $\Z\rightarrow\R$
using Hilbert's $\varepsilon$-operator (provided in Coq
standard library in module \texttt{Epsilon}), as follows:
\begin{equation}
  \dotp{f}{g} \eqdef \varepsilon \left(
  \lambda x. 
  \exists a\, b, 
  \begin{array}[t]{l}
    (\forall i, (f(i)\not=0 \lor g(i)\not=0)
    \Rightarrow a\le i \le b) \\
    \land~ x=\sum_{i=a}^b f(i)g(i)
  \end{array}
  \right)
  \label{eq:dotp2}
\end{equation}
Said otherwise, we give $\dotp{f}{g}$ a definition as a finite sum
whenever $f$ and $g$ both have finite support and we let $\dotp{f}{g}$
undefined otherwise.

To ease the manipulation of functions with finite support, we
introduce the following predicate characterizing such functions 
\begin{displaymath}
  \FS{f} \eqdef \exists a\, b, \forall i, f(i)\not=0 \Rightarrow a\le i\le b
\end{displaymath}
and we prove several lemmas about it, such as
\begin{displaymath}
  \begin{array}{l}
    \forall f g, \FS{f} \Rightarrow \FS{g} \Rightarrow \FS{f+g} \\[0.3em]
    \forall f c, \FS{f} \Rightarrow \FS{c\cdot f} \\[0.3em]
    \forall f k, \FS{f} \Rightarrow \FS{i\mapsto f(i+k)}
  \end{array}
\end{displaymath}
We also provide a Coq tactic to automatically discharge most goals
about $\FS{.}$. Finally, we can establish lemmas about the dot
product, provided functions have finite support. Here are some of
these lemmas:
\begin{displaymath}
  \begin{array}{l}
     \forall f\, g \, c, \FS{f} \Rightarrow  \FS{g} \Rightarrow 
       \dotp{c\cdot f}{g} = c\cdot\dotp{f}{g} 
     \\[0.5em]
     \forall f_1\, f_2\, g, \FS{f_1} \Rightarrow \FS{f_2} \Rightarrow 
       \FS{g} \Rightarrow 
       \dotp{f_1+f_2}{g} = \dotp{f_1}{g} + \dotp{f_2}{g} 
     \\[0.5em]
     \forall f\, g, \FS{f} \Rightarrow \FS{g} \Rightarrow 
       |\dotp{f}{g}| \le \norm{f}\cdot\norm{g} 
     \\[0.5em]
     \forall f\, g, \FS{f} \Rightarrow \FS{g} \Rightarrow 
       \norm{f+g} \le \norm{f} + \norm{g}  \\
  \end{array}
\end{displaymath}
These lemmas are proved by reduction to finite sums, thanks to
Formula~(\ref{eq:dotp2}).
Note that the value of $\prodscal{\Delta x}{f}{g}$ defined in Section~\ref{sec:discrete}
is equal to $\Delta x \cdot \dotp{f}{g}$.

\subsection{Big O notation}
\label{sec:o}

For two functions $f$ and $g$ over $\R^n$, one usually writes $f(\vec x)
= O_{\norm{\vec x} \to 0}(g(\vec x))$ for
\[\exists \alpha, C > 0,\quad \forall \vec x \in \R^n,\quad \norm{\vec x}
\le \alpha \Rightarrow |f(\vec x)| \le C \cdot |g(\vec x)|.\]

Unfortunately, this definition is not sufficient for our formalism.
Indeed, while $f(\bfx,\bfdeltax)$ will be defined over $\R^2 \times
\R^2$, $g(\bfdeltax)$ will be defined over $\R^2$ only. So it begs the
question: what to do about $\bfx$?

Our first approach was to use
\[\forall \bfx,\quad f(\bfx,\bfdeltax) = O_{\norm{\bfdeltax} \to 0}(g(\bfdeltax))\]
that is to say
\[\forall \bfx, \exists \alpha, C > 0,\quad \forall \bfdeltax \in
\R^2, \quad \norm{\bfdeltax} \le \alpha \Rightarrow
|f(\bfx,\bfdeltax)| \le C \cdot |g(\bfdeltax)| \]
which means that $\alpha$ and $C$ are functions of $\bfx$. So we would
need to take the minimum of all the possible values of $\alpha$, and the
maximum for $C$. Potentially, they may be $0$ and $+\infty$ respectively,
making them useless.

In order to solve this issue, we had to define a notion of big O uniform with respect to the additional variable $\bfx$:
\[\exists \alpha, C > 0,\quad \forall \bfx,\bfdeltax,\quad \norm{\bfdeltax}
\le \alpha \Rightarrow |f(\bfx,\bfdeltax)| \le C \cdot |g(\bfdeltax)|.\]

Variables $\bfx$ and $\bfdeltax$ are restricted to subsets $S$ and $P$
of $\R^2$. For instance, the big O that appears in
Equation~(\ref{e:convergence}) uses 
\begin{eqnarray*}
S &=& \R \times [0,\tmax],\\
P &=& \left\{ \bfdeltax = (\Delta x,\Delta t)\ |\ 0 < \Delta x \ \land\ 0 <
\Delta t \ \land\ \zeta \le \frac{c \cdot \Delta t}{\Delta x} \le 1-\xi\right\}.
\end{eqnarray*}

As often, the formal specification has allowed us to detect some flaws
in usual mathematical pen-and-paper proofs, such as an erroneous switching of
the universal and existential quantifiers hidden in the big O definition.

\subsection{Taylor expansion}
\label{sec:taylor}

The formalization assumes that ``sufficiently regular'' functions can be
uniformly approximated by multivariate Taylor series. More precisely, the
development starts by assuming that there exists two operators
\texttt{partial\_derive\_firstvar} and \texttt{\_secondvar}. Given a
real-valued function $f$ defined on the 2D plane and a point of it, they
respectively return the functions $\frac{\partial f}{\partial x}$ and
$\frac{\partial f}{\partial t}$ for this point, if they exist.

Again, these operators are similar to the use of Hilbert's $\eps$ operator. For
documentation purpose, one could add two axioms stating that the returned function
computes the derivatives for derivable functions; they are not needed
for the later development though. Indeed, none of our proofs depend on
the actual properties of derivatives; they only care about the fact that
differential operators appear in both the regularity definition below and
the wave equation.

The two primitive operators $\frac{\partial}{\partial x}$ and
$\frac{\partial}{\partial t}$ are encompassed in a generalized
differential operator $\frac{\partial^{m+n}}{\partial x^m \partial
t^n}$. This allows us to define the 2D Taylor expansion of order~$n$ of a
function~$f$:
\[\mathrm{DL}_n(f,\bfx) \eqdef (\Delta x,\Delta t) \mapsto
\sum_{p=0}^n \frac{1}{p!} \left( \sum_{m=0}^p
\binom{p}{m} \cdot \frac{\partial^p f}{\partial x^m \partial t^{p-m}}(\bfx)
\cdot \Delta x^m \cdot \Delta t^{p-m} \right).\]

\noindent
A function $f$ is then said to be sufficiently regular of order $n$ if
\begin{equation}
\forall m \le n, \quad \mathrm{DL}_{m-1}(f,\bfx)(\bfdeltax) -
f(\bfx + \bfdeltax) = O\left(\norm{\bfdeltax}^m\right).
\label{eq:taylor}
\end{equation}




%% file: wave_proof.tex
\section{The Coq Formalization: Convergence}
\label{sec:wave_proof}

\subsection{Wave equation}

As explained in Section~\ref{sec:wave}, a solution of the wave
equation with given $u_0$, $u_1$ and~$s$ verifies
Equations~(\ref{e:L})--(\ref{e:L0}).  Its discrete approximation verifies
Equations~(\ref{e:Lh})--(\ref{e:L0h}). Both are directly translated in
Coq using the definitions of Section~\ref{sec:library}.  Concerning
the discretization, we choose that the space index is in $\Z$ (to be
coherent with the dot product definition of
Section~\ref{sec:dot_prod}) while the time index is in $\N$.

Our goal is to prove the uniform convergence of the scheme with order
(2,2) on the interval $[0,\tmax]$:
\[
  \ndx{e_h^{k_{\Delta t}(t)}} =
  O_{\scriptsize
  \begin{array}{|l}
    t \in [0,\tmax] \\[1.5ex]
    (\Delta x, \Delta t) \rightarrow 0 \\
    0 < \Delta x \, \wedge \, 0 < \Delta t \, \wedge  \\
    \zeta \le c \frac{\Delta t}{\Delta x} \le 1 - \xi
  \end{array}}%
  (\Delta x^2 + \Delta t^2).\]

\subsection{Finite support}
\label{sec:FS}

The proofs concerning the convergence of the scheme rely on the dot
product. As explained in Section~\ref{sec:dot_prod}, the dot product
requires the functions to have a finite support in order to apply any
lemma. We therefore proved the finiteness of the support of many
functions. We assume that the inputs $u_0$,
$u_1$, and $s$ of the wave equation have a finite support. More
precisely, we assume that there exists $\chi_1$ and $\chi_2$ such that
$u_0(x)=u_1(x)=0$ for all $x$ out of $[\chi_1,\chi_2]$ and $s(x,t)=0$ for all
$x$ out of $[\chi_1-c \cdot t, \chi_2+c \cdot t]$ where $c$ is the velocity of
propagation of waves in Equation~(\ref{e:L}).

Figure~\ref{fig:FS} describes the nullity, that is to say the finite
support, of the various functions. We needed to prove the finiteness
of their support:
\begin{figure}[bht]
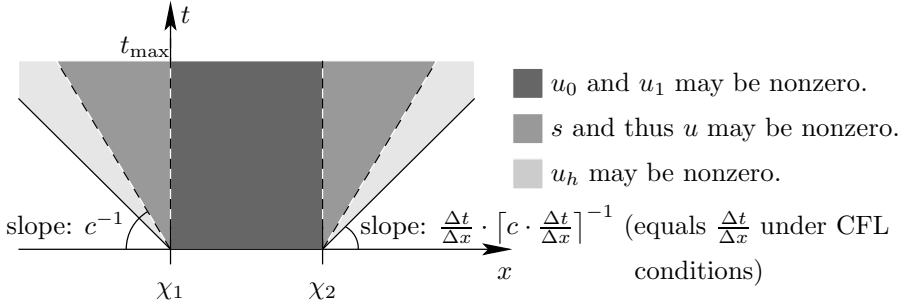
\caption{Finite supports.
  The support of the Cauchy data~$u_0$ and~$u_1$ is included in the support of
  the continuous source term~$s$, and of the continuous solution~$u$.
  Which is in turn also included in the support of the discrete solution~$u_h$,
  provided that the CFL condition holds.
  For a finite~$\tmax$, all these supports are finite.}
\label{fig:FS}
\end{figure}
\begin{itemize}
\item $u_0$ and $u_1$ by hypothesis and therefore $u_{0,j}$ and $u_{1,j}$.
\item $s$ (for any value $t$) by hypothesis and therefore $s_j^k$ is
  zero outside of a cone of slope $c^{-1}$.
\item the scheme itself has a finite support: due to the definition of
  $u_j^k$ and the nullity of $u_{0,j}$ and $u_{1,j}$ and $s_j^k$, we
  can prove that $u_j^k$ is zero outside of a cone of slope 
$\frac{\Delta t}{\Delta x} \cdot {\ceiling{c \cdot \frac{\Delta
        t}{\Delta x}}}^{-1}$. Under CFL$(\zeta, \xi)$ conditions, this 
  slope will be $\frac{\Delta t}{\Delta x}$. 
\item the truncation and convergence errors also have finite support
  with the previous slope.
\end{itemize}

We need here an axiom about the nullity of the continuous solution. We
assume that the continuous solution $u(x,t)$ is zero for $x$ out of  $[\chi_1-c \cdot
t, \chi_2+c \cdot t]$ (same as $s$). This is mathematically correct, since it
derives from d'Alembert's formula~(\ref{e:dAlembert}). But its proof is out of
the scope of the current formalization and we therefore preferred to
simply add the nullity axiom.

\subsection{Consistency}
\label{sec:consist}

We first prove that the truncation error is of order $\Delta
x^2+\Delta t^2$.  The idea is to show that, for $\bfdeltax$ small
enough, the values of the scheme $L_h$ are near the corresponding
values of $L$. This is done using the properties of Taylor
expansions. This involves long and complex expressions but the proof
is straightforward.

We first prove that the truncation error in one point $(j,k)$ is a
$O(\Delta x^2+\Delta t^2)$. This is proved for $k=0$ and $k=1$ by
taking advantage of the initializations and Taylor expansions. For
bigger $k$, the truncation error reduces to the sum of two Taylor
expansions of degree~3 in time (this means $m=4$ in
Formula~(\ref{eq:taylor})) and two Taylor expansions of degree~3 in
space that partially cancel (divided by something proportional to
$\|\bfdeltax\|^2$). Here, we take advantage of the generality of big O
as we consider the sum of a Taylor expansion on $\Delta x$ and of a
Taylor expansion on $-\Delta x$. If we had required $0 < \Delta x$ (as
a space grid step), we could not have done this proof.

The most interesting part is to go from pointwise consistency to
uniform consistency. We want to prove that the norm of the
truncation error (in the sense of the infinite dot product $\psdx{\cdot}{\cdot}$)
is also $O(\Delta x^2+\Delta t^2)$. We therefore need to bound the number of
nonzero values of the truncation error. As explained in
Section~\ref{sec:FS}, the truncation error values at time $k
\cdot \Delta t$ may be nonzero between
$ {\chi_1}'_k = \left\lfloor \frac{\chi_1}{\Delta x} \right\rfloor 
  - \left\lceil c \cdot \frac{\Delta t}{\Delta x} \right\rceil k$ and
${\chi_2}'_k = \left\lceil \frac{\chi_2}{\Delta x}  \right\rceil
  + \left\lceil c \cdot \frac{\Delta t}{\Delta x} \right\rceil
k$. This gives a number of terms $N$ roughly bounded by (all
details are handled in the formal proof):
\begin{eqnarray*}
N & \le & \frac{{\chi_2}'_k-{\chi_1}'_k}{\Delta x} \le \frac{\chi_2-\chi_1}{\Delta x^2} +2 \cdot k_{\max} 
\cdot \frac{\ceiling{c \cdot \frac{\Delta t}{\Delta x}}}{\Delta x} \\ 
&\le& \frac{\chi_2-\chi_1}{\Delta x^2} +2 \cdot \frac{\tmax}{\Delta
t} \cdot \frac{c \cdot \frac{\Delta t}{\Delta x} +1}{\Delta x} \\
%
\end{eqnarray*}
As the norm is a $\Delta x$-norm, this reduces to bounding with a
constant value the value $N \cdot \Delta x^2$ which is smaller than
$\chi_2-\chi_1 + 2 \cdot \tmax \cdot c + 2 \cdot \tmax \cdot \frac{\Delta x}{\Delta
t}$. To bound this with a constant value, we require $c \frac{\Delta
t}{\Delta x}$ to have a constant lower bound $\zeta$ (it already had an
upper bound $1-\xi$). Then $N \cdot \Delta x^2 \le \chi_2-\chi_1 +
2 \cdot \tmax \cdot c + 2 \cdot c \cdot \tmax \cdot \frac{1}{\zeta}$ which
is constant.

\begin{figure}[th]
\centerline{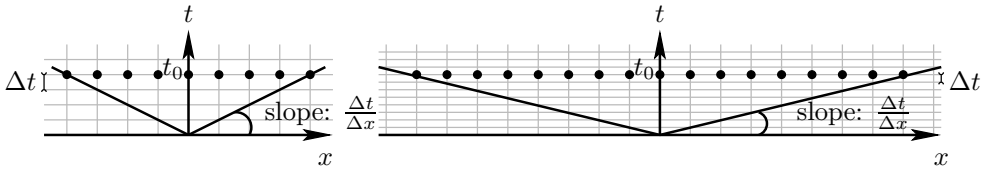}
\caption{For a given time $t_0$, the number of nonzero values increases
when the slope $\frac{\Delta t}{\Delta x}$ goes to zero.
From left to right, $\Delta t$ is divided by~2 whereas~$\Delta x$ remains the
same.
We can see that the number of nonzero terms is almost doubled (from~9 to~17).}
\label{fig:eta}
\end{figure}

Mathematically, this requirement comes as a surprise. The following scenario explains it. If $c \frac{\Delta t}{\Delta x}$ goes to zero,
then $\Delta t$ goes to zero much faster than $\Delta x$. It
corresponds to Figure~\ref{fig:eta}. The number of nonzero terms (for
$u_h$ and thus for the truncation error) goes to infinity as 
$\frac{\Delta t}{\Delta x}$ goes to zero.

\subsection{Stability}
\label{sec:stab}

To prove stability, we use the discrete energy defined in Equation~(\ref{e:discreteenergy}). From the properties of the scheme, we calculate the evolution of the
energy. At each step, it increases by a known value. In particular, if
$s$ is zero, the discrete energy (as the continuous energy) is constant:
$$ \forall k > 0, \quad
    E_h (c) (u_h)^{k + \demi} - E_h (c) (u_h)^{k - \demi} =
    \demi \prodscal{\Delta x}{u_h^{k + 1} - u_h^{k - 1}}{s_h^k}.
$$

\noindent
From this, we give an underestimation of the energy:
$$  \forall k, \quad
   \demi\left(1-\left(c \frac{\Delta t}{\Delta x}\right)^2\right)
  \norme{\Delta x}{\frac{u_h^{k + 1} - u_h^k}{\Delta t}} 
  \le {E_h (c) (u_h)^{k + \demi}}.$$
Therefore we have the nonnegativity of the
energy under CFL$(\zeta,\xi)$ conditions.
For convergence, the key result is the overestimation of the energy:
$$\sqrt{E_h (c) (u_h)^{k + \demi}} \le
\sqrt{E_h (c) (u_h)^{\demi}} + \frac{\sqrt{2}}{2\sqrt{2 \xi -
    \xi^2}} \cdot \Delta t \cdot \sum_{j=1}^k 
    \norme{\Delta x}{i \mapsto s_h(i,j)}$$
for all time $t$, with $k=\left\lfloor
\frac{t}{\Delta t}\right\rfloor -1$.

This completes the stability proof. In the inequality above, the energy is
bounded for $u_h$, but the bound is actually valid for all the solutions
of the discrete scheme, for any initial conditions and source term.

Note that the formal proof of stability closely follows the mathematical
pen-and-paper proof and no additional hypotheses were found to be
necessary.

\subsection{Convergence}

We prove that the convergence error is the solution
of a scheme and therefore the results of Section~\ref{sec:stab} apply
to it. More precisely, for all $\bfdeltax$, the convergence error is
solution of a discrete scheme with inputs
$$u_{0,j}=0, \qquad \qquad u_{1,j}=\frac{e_j^1}{\Delta t}, \qquad \mbox{and}
\qquad s_j^k= \eps_j^{k+1},$$
where the errors refer to the errors of the initial scheme
of the wave equation with grid steps $\bfdeltax$. (Actual Coq notations depend
on many more variables.)

We have proved many lemmas about the initializations of our
scheme and of the convergence error. The idea is to prove that the
initializations of the scheme are precise enough to guarantee that
the initial convergence errors (at step 0 and 1) are accurate enough. 

We also bounded the energy of the convergence error. Using
results of Section~\ref{sec:stab}, the proof reduces to bounding the sum of
the source terms, here the truncation errors. Using results of
Section~\ref{sec:consist}, we prove this sum to be $O(\Delta x^2+\Delta
t^2)$. A few more steps conclude the proof.

Once more, the formal proof follows the pen-and-paper proof and progresses smoothly under the required
hypothesis, including all the conditions on $\frac{\Delta t}{\Delta
  x}$ of Equation (\ref{eq:cfl}).

%% file: FS.pdf_t
\begin{picture}(0,0)%
\includegraphics{FS.eps}%
\end{picture}%
\setlength{\unitlength}{4144sp}%
\begingroup\makeatletter\ifx\SetFigFont\undefined%
\gdef\SetFigFont#1#2#3#4#5{%
  \reset@font\fontsize{#1}{#2pt}%
  \fontfamily{#3}\fontseries{#4}\fontshape{#5}%
  \selectfont}%
\fi\endgroup%
\begin{picture}(4395,1831)(1114,-2870)
\put(1756,-2446){\makebox(0,0)[rb]{\smash{{\SetFigFont{10}{12.0}{\familydefault}{\mddefault}{\updefault}{\color[rgb]{0,0,0}slope: $c^{-1}$}%
}}}}
\put(2026,-2806){\makebox(0,0)[b]{\smash{{\SetFigFont{10}{12.0}{\familydefault}{\mddefault}{\updefault}{\color[rgb]{0,0,0}$\chi_1$}%
}}}}
\put(2926,-2806){\makebox(0,0)[b]{\smash{{\SetFigFont{10}{12.0}{\familydefault}{\mddefault}{\updefault}{\color[rgb]{0,0,0}$\chi_2$}%
}}}}
\put(4276,-2131){\makebox(0,0)[lb]{\smash{{\SetFigFont{10}{12.0}{\familydefault}{\mddefault}{\updefault}{\color[rgb]{0,0,0}$u_h$ may be nonzero.}%
}}}}
\put(4276,-1861){\makebox(0,0)[lb]{\smash{{\SetFigFont{10}{12.0}{\familydefault}{\mddefault}{\updefault}{\color[rgb]{0,0,0}$s$ and thus $u$ may be nonzero.}%
}}}}
\put(4276,-1591){\makebox(0,0)[lb]{\smash{{\SetFigFont{10}{12.0}{\familydefault}{\mddefault}{\updefault}{\color[rgb]{0,0,0}$u_0$ and $u_1$ may be nonzero.}%
}}}}
\put(3151,-2446){\makebox(0,0)[lb]{\smash{{\SetFigFont{10}{12.0}{\familydefault}{\mddefault}{\updefault}{\color[rgb]{0,0,0}slope: $\frac{\Delta t}{\Delta x} \cdot \left\lceil c \cdot \frac{\Delta t}{\Delta x} \right\rceil^{-1}$ (equals $\frac{\Delta t}{\Delta x}$ under CFL}%
}}}}
\put(4006,-2716){\makebox(0,0)[b]{\smash{{\SetFigFont{10}{12.0}{\familydefault}{\mddefault}{\updefault}{\color[rgb]{0,0,0}$x$}%
}}}}
\put(2116,-1186){\makebox(0,0)[b]{\smash{{\SetFigFont{10}{12.0}{\familydefault}{\mddefault}{\updefault}{\color[rgb]{0,0,0}$t$}%
}}}}
\put(2026,-1366){\makebox(0,0)[rb]{\smash{{\SetFigFont{10}{12.0}{\familydefault}{\mddefault}{\updefault}{\color[rgb]{0,0,0}$\tmax$}%
}}}}
\put(4771,-2716){\makebox(0,0)[lb]{\smash{{\SetFigFont{10}{12.0}{\familydefault}{\mddefault}{\updefault}{\color[rgb]{0,0,0}conditions)}%
}}}}
\end{picture}%

%% file: eta.pdf_t
\begin{picture}(0,0)%
\includegraphics{eta.eps}%
\end{picture}%
\setlength{\unitlength}{4144sp}%
\begingroup\makeatletter\ifx\SetFigFont\undefined%
\gdef\SetFigFont#1#2#3#4#5{%
  \reset@font\fontsize{#1}{#2pt}%
  \fontfamily{#3}\fontseries{#4}\fontshape{#5}%
  \selectfont}%
\fi\endgroup%
\begin{picture}(5430,1033)(-464,-3986)
\put(4906,-3931){\makebox(0,0)[b]{\smash{{\SetFigFont{10}{12.0}{\familydefault}{\mddefault}{\updefault}{\color[rgb]{0,0,0}$x$}%
}}}}
\put(1261,-3931){\makebox(0,0)[b]{\smash{{\SetFigFont{10}{12.0}{\familydefault}{\mddefault}{\updefault}{\color[rgb]{0,0,0}$x$}%
}}}}
\put(-449,-3481){\makebox(0,0)[rb]{\smash{{\SetFigFont{10}{12.0}{\familydefault}{\mddefault}{\updefault}{\color[rgb]{0,0,0}$\Delta t$}%
}}}}
\put(4951,-3481){\makebox(0,0)[lb]{\smash{{\SetFigFont{10}{12.0}{\familydefault}{\mddefault}{\updefault}{\color[rgb]{0,0,0}$\Delta t$}%
}}}}
\put(451,-3076){\makebox(0,0)[b]{\smash{{\SetFigFont{10}{12.0}{\familydefault}{\mddefault}{\updefault}{\color[rgb]{0,0,0}$t$}%
}}}}
\put(3241,-3076){\makebox(0,0)[b]{\smash{{\SetFigFont{10}{12.0}{\familydefault}{\mddefault}{\updefault}{\color[rgb]{0,0,0}$t$}%
}}}}
\put(361,-3391){\makebox(0,0)[b]{\smash{{\SetFigFont{10}{12.0}{\familydefault}{\mddefault}{\updefault}{\color[rgb]{0,0,0}$t_0$}%
}}}}
\put(3151,-3391){\makebox(0,0)[b]{\smash{{\SetFigFont{10}{12.0}{\familydefault}{\mddefault}{\updefault}{\color[rgb]{0,0,0}$t_0$}%
}}}}
\put(901,-3661){\makebox(0,0)[lb]{\smash{{\SetFigFont{10}{12.0}{\familydefault}{\mddefault}{\updefault}{\color[rgb]{0,0,0}slope: $\frac{\Delta t}{\Delta x}$}%
}}}}
\put(4051,-3661){\makebox(0,0)[lb]{\smash{{\SetFigFont{10}{12.0}{\familydefault}{\mddefault}{\updefault}{\color[rgb]{0,0,0}slope: $\frac{\Delta t}{\Delta x}$}%
}}}}
\end{picture}%

%% file: conclusion.tex
\section{Conclusion and perspectives}
\label{sec:conclusion}

One of the goals of this work is to favor the use of formal methods in
numerical analysis. It may seem to be just wishful thinking, but it is
actually seen as needed by some applied mathematicians. An early case led to
the certification of the O$\partial$yss\'ee tool~\cite{May02}. This tool
performs automatic differentiation, which is one of the basic blocks for
\emph{gradient}-based algorithms. Our work tackles the converse problem:
instead of considering derivation-based algorithms, we have formalized
and proved part of the mathematical background behind integration-based
algorithms.

This work shows there may be a synergy between applied mathematicians
and logicians. Both domains are required here: applied mathematics for
an initial proof that could be enriched upon request and formal
methods for machine-checking it. This may be the reason why such
proofs are scarce as this kind of collaboration is uncommon.

Proof assistants seem to mainly deal with algebra, but we have
demonstrated that formalizing numerical analysis is possible too. We can
confirm that pen-and-paper proofs are sometimes sketchy: they may be
fuzzy about the needed hypotheses, especially when switching
quantifiers. We have also learned that filling the gaps may cause us to
go back to the drawing board and to change the basic blocks of our
formalization to make them more generic (a big O that needs to be
uniform and also generic with respect to a property $P$).

The formal bound on the error method, while of mathematical interest, is
not sufficient to guarantee the correction of numerical applications
implementing the three-point scheme. Indeed, such applications usually
perform approximated computations, \emph{e.g.}, floating-point
computations, for efficiency and simplicity reasons. As a consequence,
the proof of the method error has to be combined with a proof on the
rounding error, in order to get a full-fledged correction proof.
Fortunately, the proof on the rounding error has already been
achieved~\cite{Bol09}. We are therefore close to having a formal proof of
both the numerical scheme and its floating-point implementation.

An advantage of Coq with respect to most other proof assistants is the
ability to \emph{extract} programs from proofs~\cite{Let02}. For this
work, it does not make much sense to extract the algorithm from the
proofs: not only is the algorithm already well-known, but its
floating-point implementation was also certified~\cite{Bol09}. So, an
extraction of the algorithm would not bring much. However, extraction
gives access to the constant $C$ hidden behind the big O
notation. Indeed, the proof of the floating-point algorithm relies on
the discrete solution being good enough, so that the computed result
does not diverge. Precisely, the convergence error has to be smaller
than~$1$, and an extracted computation would be able to ensure this
property. Furthermore, having access to this constant can be useful to
the applied mathematicians for the a posteriori estimations needed for
adaptive mesh refinements. Extraction also gives access to the
$\alpha$ constant. That way, we could check that the constant $\bfdeltax$
chosen in the C program described in \cite{Bol09} verifies this
requirement. Note that performing an extraction requires to modify the
definition of the big O so that it lives in \texttt{Set} instead of
\texttt{Prop}. But this formalization change happens to be
straightforward and Coq then succeeds in extracting mathematical formulas
for constants $\alpha$ and $C$. Only basic operators (\emph{e.g.} $+$,
$\sqrt{\cdot}$, $\min$) and constants (\emph{e.g.} $\tmax$,
$\xi$, $\chi_1$, Taylor constants) appear in them, so they should
be usable in practice.

\begin{figure}[htb]
\centerline{\includegraphics{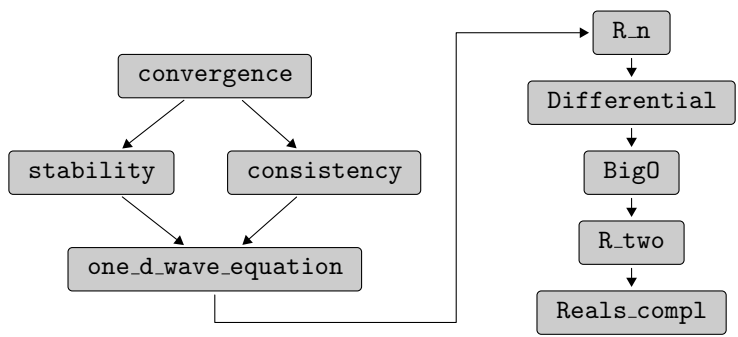}}
\caption{Dependency graph of the Coq development. On the left are the
  files from the convergence proof. The other
  files correspond to the reusable library. }
\label{fig:dep}
\end{figure}

The formal development is about 4500-line long. Its dependency graph is
detailed in Figure~\ref{fig:dep}. About half of the development is a
reusable library described in Section~\ref{sec:library} and the other
half is the proof of convergence of the numerical scheme described in
Section~\ref{sec:wave_proof}. This may seem a long proof for a
single scheme for a single PDE. To put it into perspective, usual
pen-and-paper proofs are 10-page long and an in-depth proof can be
60-page long. (We wrote one to ensure that we were not getting
sidetracked.) So, at least from a length point of view, the formal proof
is comparable to a detailed pen-and-paper proof.

In the end, the whole development contains only two axioms: the
$\varepsilon$ operator for the infinite dot product (see
Section~\ref{sec:dot_prod}) and the finite support of the
continuous solution of the wave equation (see Section~\ref{sec:FS}). So,
except for this last axiom which is related to the chosen PDE, the
full numerical analysis proof of convergence is machine-checked and
all required hypotheses are made clear. There is no loss of confidence
due to this axiom, since the kind of proof and the results it is based upon are completely
different from the ones presented here. Indeed, this axiom is about continuous
solutions and hence much less error-prone.

For this exploratory work, we only considered the simple three-point scheme for
the one-dimensional wave equation. Further works involve generalizing our
approach to other
schemes and other PDEs. We are confident that it would scale to higher-dimension
and higher-order equations solved by discrete numerical schemes. However, the
proofs of Section~\ref{sec:wave_proof} are entangled with particulars of the
presented problem, and would therefore have to be redone for other problems. So
a more fruitful approach would be to prove once and for all the Lax equivalence
theorem that states that consistency implies the equivalence between
convergence and stability. This would considerably reduce the amount of work
needed for tackling other schemes and equations.

This work also showed us that summations and finite support functions
play a much more important role in the development than we first
expected. We are therefore considering moving to the SSReflect interface
and libraries for Coq~\cite{BGOP08}, so as to simplify the manipulations of
these objects in our forthcoming works.




%% file: RR-7181.bbl
\begin{thebibliography}{10}

\bibitem{ach:wpe:73}
J.~D. Achenbach.
\newblock {\em {Wave Propagation in Elastic Solids}}.
\newblock North Holland, Amsterdam, 1973.

\bibitem{AD07}
Jeremy Avigad and Kevin Donnelly.
\newblock {A Decision Procedure for Linear "Big O" Equations}.
\newblock {\em J. Autom. Reason.}, 38(4):353--373, 2007.

\bibitem{bec:esn:09}
\'E. B\'ecache.
\newblock {\'E}tude de sch\'emas num\'eriques pour la r\'esolution de
  l'\'equation des ondes.
\newblock Master Mod\'elisation et simulation, Cours ENSTA,
  \url{http://www-rocq.inria.fr/~becache/COURS-ONDES/Poly-num-0209.pdf}, 2009.

\bibitem{Coq}
Yves Bertot and Pierre Cast\'eran.
\newblock {\em {Interactive Theorem Proving and Program Development. Coq'Art:
  The Calculus of Inductive Constructions}}.
\newblock Texts in Theoretical Computer Science. Springer, 2004.

\bibitem{BGOP08}
Yves Bertot, Georges Gonthier, Sidi Ould~Biha, and Ioana Pasca.
\newblock {Canonical Big Operators}.
\newblock In {\em 21st International Conference on Theorem Proving in Higher
  Order Logics (TPHOLs'08)}, volume 5170 of {\em LNCS}, pages 86--101,
  Montreal, Canada, 2008. Springer.

\bibitem{Bol09}
Sylvie Boldo.
\newblock {Floats \& {R}opes: a case study for formal numerical program
  verification}.
\newblock In {\em Proceedings of the 36th International Colloquium on Automata,
  Languages and Programming}, volume 5556 of {\em LNCS}, pages 91--102, Rhodos,
  Greece, 2009. Springer.

\bibitem{bg:mcw:94}
L.~M. Brekhovskikh and V.~Goncharov.
\newblock {\em {Mechanics of Continua and Wave Dynamics}}.
\newblock Springer, 1994.

\bibitem{cfl:pde:67}
R.~Courant, K.~Friedrichs, and H.~Lewy.
\newblock On the partial difference equations of mathematical physics.
\newblock {\em IBM Journal of Research and Development}, 11(2):215--234, 1967.

\bibitem{Cru02}
Lu{\'i}s Cruz-Filipe.
\newblock {A Constructive Formalization of the Fundamental Theorem of
  Calculus}.
\newblock In Herman Geuvers and Freek Wiedijk, editors, {\em Proceedings of the
  2nd International Workshop on Types for Proofs and Programs (TYPES 2002)},
  volume 2646 of {\em LNCS}, Berg en Dal, Netherlands, 2002. Springer.

\bibitem{Dut96}
Bruno Dutertre.
\newblock {Elements of Mathematical Analysis in {PVS}}.
\newblock In Joakim von Wright, Jim Grundy, and John Harrison, editors, {\em
  Proceedings of the 9th International Conference on Theorem Proving in Higher
  Order Logics (TPHOLs'96)}, volume 1125 of {\em LNCS}, pages 141--156, Turku,
  Finland, 1996. Springer.

\bibitem{Fle00}
Jacques~D. Fleuriot.
\newblock {On the Mechanization of Real Analysis in Isabelle/HOL}.
\newblock In Mark Aagaard and John Harrison, editors, {\em 13th International
  Conference on Theorem Proving and Higher-Order Logic (TPHOLs'00)}, volume
  1869 of {\em LNCS}, pages 145--161. Springer, 2000.

\bibitem{GK01}
Ruben Gamboa and Matt Kaufmann.
\newblock {Nonstandard Analysis in ACL2}.
\newblock {\em Journal of Automated Reasoning}, 27(4):323--351, 2001.

\bibitem{NiqGeu00}
Herman Geuvers and Milad Niqui.
\newblock {Constructive Reals in {C}oq: Axioms and Categoricity}.
\newblock In Paul Callaghan, Zhaohui Luo, James McKinna, and Robert Pollack,
  editors, {\em Proceedings of the 1st International Workshop on Types for
  Proofs and Programs (TYPES 2000)}, volume 2277 of {\em LNCS}, pages 79--95,
  Durham, United Kingdom, 2002. Springer.

\bibitem{Har98}
John Harrison.
\newblock {\em {Theorem Proving with the Real Numbers}}.
\newblock Springer, 1998.

\bibitem{Har05}
John Harrison.
\newblock {A HOL Theory of Euclidean Space}.
\newblock In Joe Hurd and Thomas~F. Melham, editors, {\em 18th International
  Conference on Theorem Proving and Higher-Order Logic (TPHOLs'05)}, volume
  3603 of {\em LNCS}, pages 114--129. Springer, 2005.

\bibitem{joh:pde:86}
F.~John.
\newblock {\em {Partial Differential Equations}}.
\newblock Springer, 1986.

\bibitem{dal:rcf:47}
J.~{le Rond D'Alembert}.
\newblock Recherches sur la courbe que forme une corde tendue mise en
  vibrations.
\newblock In {\em Histoire de l'Acad\'emie Royale des Sciences et Belles
  Lettres (Ann\'ee 1747)}, volume~3, pages 214--249. Haude et Spener, Berlin,
  1749.

\bibitem{Let02}
Pierre Letouzey.
\newblock {A New Extraction for {Coq}}.
\newblock In Herman Geuvers and Freek Wiedijk, editors, {\em Proceedings of the
  2nd International Workshop on Types for Proofs and Programs (TYPES 2002)},
  volume 2646 of {\em LNCS}, Berg en Dal, Netherlands, 2003. Springer.

\bibitem{May01}
Micaela Mayero.
\newblock {\em Formalisation et automatisation de preuves en analyses r\'eelle
  et num\'erique}.
\newblock PhD thesis, Universit\'e Paris VI, 2001.

\bibitem{May02}
Micaela Mayero.
\newblock {Using Theorem Proving for Numerical Analysis (Correctness Proof of
  an Automatic Differentiation Algorithm)}.
\newblock In Victor Carre{\~n}o, C{\'e}sar Mu{\~n}oz, and Sofi{\`e}ne Tahar,
  editors, {\em 15th International Conference on Theorem Proving and
  Higher-Order Logic}, volume 2410 of {\em LNCS}, pages 246--262, Hampton, VA,
  USA, 2002. Springer.

\bibitem{new:alm:87}
I.~Newton.
\newblock {Axiomata, sive Leges Motus}.
\newblock In {\em Philosophiae Naturalis Principia Mathematica}, volume~1.
  London, 1687.

\bibitem{tho:npd:95}
James~William Thomas.
\newblock {\em {Numerical Partial Differential Equations: Finite Difference
  Methods}}.
\newblock Number~22 in Texts in Applied Mathematics. Springer, 1995.

\bibitem{zwi:hde:98}
D.~Zwillinger.
\newblock {\em {Handbook of Differential Equations}}.
\newblock Academic Press, 1998.

\end{thebibliography}
